\documentclass[%
 aip,
 amsmath,amssymb,
 reprint,%
]{revtex4-2}

\usepackage{graphicx}
\usepackage{subcaption}

\usepackage{dcolumn}
\usepackage{bm}

\usepackage[utf8]{inputenc}
\usepackage[T1]{fontenc}
\usepackage{mathptmx}
\usepackage{etoolbox}

\usepackage{color}
\definecolor{gray}{rgb}{0.4, 0.4, 0.4}
\definecolor{dkred}{rgb}{0.8, 0.0, 0.0}
\definecolor{dkgreen}{rgb}{0.0, 0.5, 0.0}
\definecolor{dkblue}{rgb}{0.0, 0.0, 0.5}

\makeatletter
\def\@email#1#2{%
 \endgroup
 \patchcmd{\titleblock@produce}
  {\frontmatter@RRAPformat}
  {\frontmatter@RRAPformat{\produce@RRAP{*#1\href{mailto:#2}{#2}}}\frontmatter@RRAPformat}
  {}{}
}%
\makeatother
\begin{document}

\preprint{AIP/123-QED}

\title[940-nm VCSELs grown by molecular beam epitaxy on Ge(001)]{940-nm VCSELs grown by molecular beam epitaxy on Ge(001)}
\author{K. Ben Saddik$^*$}%
\affiliation{ 
LAAS-CNRS, Université de Toulouse, CNRS, INSA, 31400 Toulouse, France
}
\author{A. Arnoult}%

\affiliation{ 
LAAS-CNRS, Université de Toulouse, CNRS, INSA, 31400 Toulouse, France
}
\author{P. Gadras}%
\affiliation{ 
LAAS-CNRS, Université de Toulouse, CNRS, INSA, 31400 Toulouse, France
}
\author{S. Calvez}%
\affiliation{ 
LAAS-CNRS, Université de Toulouse, CNRS, INSA, 31400 Toulouse, France
}
\author{L. Bourdon}%
\affiliation{ 
LAAS-CNRS, Université de Toulouse, CNRS, INSA, 31400 Toulouse, France
}
\author{R. Monflier}%
\affiliation{ 
LAAS-CNRS, Université de Toulouse, CNRS, INSA, 31400 Toulouse, France
}
\author{W.Strupinski} 
\affiliation{ 
Vigo System S.A., Warsaw, Poland
}
\affiliation{ 
Faculty of Physics, Warsaw University of Technology, 00-662 Warszawa, Poland
}
\author{G. Almuneau$^*$}%
\email{almuneau@laas.fr}%
\affiliation{ 
LAAS-CNRS, Université de Toulouse, CNRS, INSA, 31400 Toulouse, France
}%

\date{\today}

\begin{abstract}
Vertical-cavity surface-emitting laser (VCSEL) structures emitting near 940~nm were grown by solid source molecular beam epitaxy (MBE) on Ge(001) substrates. The VCSEL MBE-growth was realized upon a virtual substrate composed of GaAs on Ge grown by melatorganic vapour phase epitaxy (MOVPE). In situ monitoring during MBE growth employed multispectral reflectometry and magnification-inferred curvature imaging for real-time growth analysis. Curvature measurements revealed progressive compressive stress, while optical reflectivity data confirmed uniform layer growth and accurate stopband formation. Fabricated devices with mesa diameters of 35–40 $\mu$m, corresponding to oxide apertures of approximately 11-16~$\mu$m, exhibited room-temperature lasing under continuous-wave bias with threshold currents below 3~mA. To the best of our knowledge, this is the first demonstration of monolithically integrated 940~nm VCSELs grown on Ge substrates by MBE. These results confirm the viability of MBE-grown VCSELs on Ge with in situ process control for scalable optoelectronic integration.
\end{abstract}

\maketitle


Vertical cavity surface emitting lasers (VCSELs) operating near  940~nm are fundamental light sources in consumer electronics, automotive sensing, and free-space optical communications~\cite{Iga2000,Yoshikawa2005,Cheng2018a}. Their compact architecture, low current threshold, and high modulation speed make them very attractive for scalable integration into photonic systems~\cite{Hofmann2012,Song2015,Lavrencik2018}. Historically, VCSELs at this wavelength are fabricated on GaAs substrates using metal-organic vapour phase epitaxy (MOVPE) or molecular beam epitaxy (MBE), where the AlGaAs/GaAs system with a compatible lattice parameter offers high-reflectivity distributed Bragg reflectors (DBRs) and efficient quantum well gain medium\cite{Chua1997,Cheng2018a,Baker2021}.
As photonic-electronic integration advances, interest has grown in transferring III–V VCSEL technology onto group IV substrates. Ge offers several advantages in this context. Its lattice constant, matching with Al$_{0.6}$Ga$_{0.4}$As composition, lies very close between those of GaAs and AlAs with mismatches below $10^{-3}$, thus minimizing the epitaxial strain in GaAs/AlGaAs DBR grown on Ge compared to GaAs substrates. 

Moreover, Ge is available in large wafer sizes compatible with CMOS back-end processes \cite{Depuydt2006}. These factors make Ge an attractive platform for monolithic VCSEL integration. Although recent reports have demonstrated partial and complete VCSEL structures on Ge, grown by MOVPE exclusively\cite{Wan2024,Zhao2024}, the ability to monitor and control epitaxial strain, optical thickness and cavity alignment during growth remains largely unexplored. Furthermore, to our knowledge, there are no reports of MBE epitaxy of AlGaAs VCSELs on germanium substrates.  

MBE offers precise control over layer thickness, composition and doping profiles. Abrupt heterointerfaces, which are required for high-reflectivity DBRs, are achieved through rapid shutter switching. Digital alloying is employed to synthesise effective alloys and engineer compositionally graded profiles via sequences of ultrathin sublayers. Additionally, MBE systems can be equipped with in situ instruments that provide direct insight into the evolution of structural and optical properties during growth. These capabilities are particularly valuable for fabricating VCSELs on Ge, as strain evolution and thermal mismatch must be carefully managed to maintain high structural quality and achieve the desired optical performance. 
A further significant advantage of using MBE for the heteroepitaxy of AlGaAs multilayer components on Ge substrates is the ability to achieve this at a lower temperature than with MOVPE, with potential benefits for the crystalline quality of III-V materials on Ge.

Here, we present the monolithic growth of 940 nm VCSELs on Ge(001) substrates by MBE, with in situ monitoring of the strain evolution and the optical cavity building up. The combined use of curvature measurement and broadband reflectometry enables real-time feedback on structural and optical parameters during the epitaxial growth. This approach is particularly relevant for the controlled fabrication of strain-sensitive III-V vertical cavity devices on group IV substrates. Room-temperature lasing was achieved from fabricated devices with threshold currents below 3 mA. These results establish the viability of MBE-grown VCSELs on Ge with integrated process diagnostics for scalable photonic integration.


The 940~nm VCSEL structures were grown on a N-doped 4-inch Ge wafer with a 6$^{\circ}$ miscut toward the (111) direction, supplied by UMICORE. In a first step, a 100~nm GaAs buffer layer was grown using MOVPE to promote high-quality nucleation and minimize defect formation at the Ge interface\cite{Kolkowski2024}. The VCSEL  structure was then entirely grown by MBE on a Riber 412 system, equipped with dual Ga and Al effusion cells to enable both continuous and digital alloying growth modes. A dedicated In cell was used to grow the quantum wells, and a valved cracker cell provided As$_4$ flux with a beam equivalent pressure (BEP) of $2.0 \times 10^{-5}$~Torr, corresponding to a V/III ratio of 3.2.

N-type doping was achieved using a Si effusion cell calibrated for a carrier concentration of approximately $2 \times 10^{18}$~cm$^{-3}$. P-type doping was introduced via carbon, using a CBr$_4$ precursor. The growth rates were 1.0~\textmu m/h for Al$_{0.9}$Ga$_{0.1}$As, and 0.5~\textmu m/h for both GaAs and digitally alloyed layers.

The full epitaxial design includes a 35-pair n-doped bottom Bragg mirror (DBR), a half-$\lambda$ cavity including 3 quantum wells, and a 17 periods p-doped top DBR. The active region consists of three 7.5~nm undoped In$_{0.1}$Ga$_{0.9}$As quantum wells, separated by 10~nm undoped GaAs barrier layers. Each DBR consists of alternating $\lambda$/4-thick layers of Al$_{0.9}$Ga$_{0.1}$As and GaAs, centered at 940~nm. Nominally, the Al$_{0.9}$Ga$_{0.1}$As layers are 56.6~nm thick, and the GaAs layers are 48.1~nm thick. Each DBR pair also includes 20~nm graded transition layers between the high and low index materials to smooth the compositional change\cite{Zhou1991,Pickrell2005,Cho2010}. Digital alloying was applied to the layers with high aluminum content, as well as to the graded transition layers that form part of each DBR pair.

Above the active region, a 30~nm Al$_{0.98}$Ga$_{0.02}$As layer was included to form a lateral oxidation aperture. The top p-doped cladding was composed of Al$_{0.4}$Ga$_{0.6}$As gradually transitioned to Al$_{0.8}$Ga$_{0.2}$As, followed by a p$^{+}$ GaAs contact cap.


The epitaxial growth was monitored \textit{in situ} using two complementary real-time diagnostics integrated into the MBE chamber. Strain evolution was tracked using the EZ-CURVE system (Riber), which employs magnification-inferred curvature (MIC) imaging to quantify changes in wafer bowing with micrometer-level sensitivity\cite{Arnoult2021,BenSaddik2025}. Data were acquired at rates up to 100~Hz, enabling the detection of both layer-by-layer strain variations and cumulative strain across the DBR stacks and the active region.

Simultaneously, broadband optical reflectometry was performed using the EZ-REF system (Riber), which captures optical reflectance spectra in real time over a wavelength range of 320~nm to 1700~nm. This allowed continuous monitoring of Fabry--Pérot interference oscillations and DBR stopband formation during growth. 
Together, these tools provided layer-resolved feedback on the mechanical and optical evolution of the structure, ensuring cavity resonance targeted wavelength.

Following growth, the wafer was characterized \textit{ex situ} to assess structural quality and optical performance. Reflectivity spectra at normal incidence were measured using a Bruker Vertex 70 Fourier-transform infrared (FTIR) spectrometer equipped with a tungsten halogen light source and an InGaAs detector. The resulting data revealed a DBR stopband centered at 942.7~nm with a full width bandwidth of 93.6~nm, in good agreement with the designed optical cavity.


VCSEL devices were fabricated using standard process flow, including mesa etching and selective lateral oxidation to define the optical and electrical aperture. Circular mesas with diameters of 35\,µm, 36\,µm, 37.5\,µm, and 40\,µm were patterned using photolithography and etched down to the bottom DBR, ensuring vertical electrical access and optical confinement. The etching was performed using dry etching ICP-RIE with gases Cl$_2$ and N$_2$. The etching depth reached up to the 5 pairs of the bottom DBR, ensuring efficient vertical electrical access. Selective oxidation was performed in a water vapor-enriched $N_{2}/H_{2}$ atmosphere at a temperature of 430°C. The Al$_{0.98}$Ga$_{0.02}$As oxidation layer formed a lateral oxide aperture with a calibrated horizontal oxidation depth of 15~µm from the mesa edge. The resulting oxide aperture diameters were approximately 5–10~µm, depending on the initial mesa size. These apertures defined the lateral optical confinement and current injection zone. The oxidation process was monitored in real time, allowing for precise control over the oxidation depth\cite{Almuneau2008}. 
Top and bottom metal contacts were deposited using electron beam evaporation. The p-contact was made of Ti~(10~nm)/Pt~(50~nm)/Au~(400~nm), while the n-contact was made of Ni~(5~nm)/AuGeNi~(250~nm)/Au~(250~nm). The devices were then annealed at 400°C during 60 sec in a N$_{2}$ atmosphere to ensure proper ohmic contact formation. Lasing performance was evaluated under continuous-wave (CW) operation at 25°C. Light–current–voltage (LIV) measurements were performed using a calibrated silicon photodiode positioned normal to the emission surface. 



\begin{figure*}[htbp]
  \centering
  \includegraphics[width=1\textwidth]{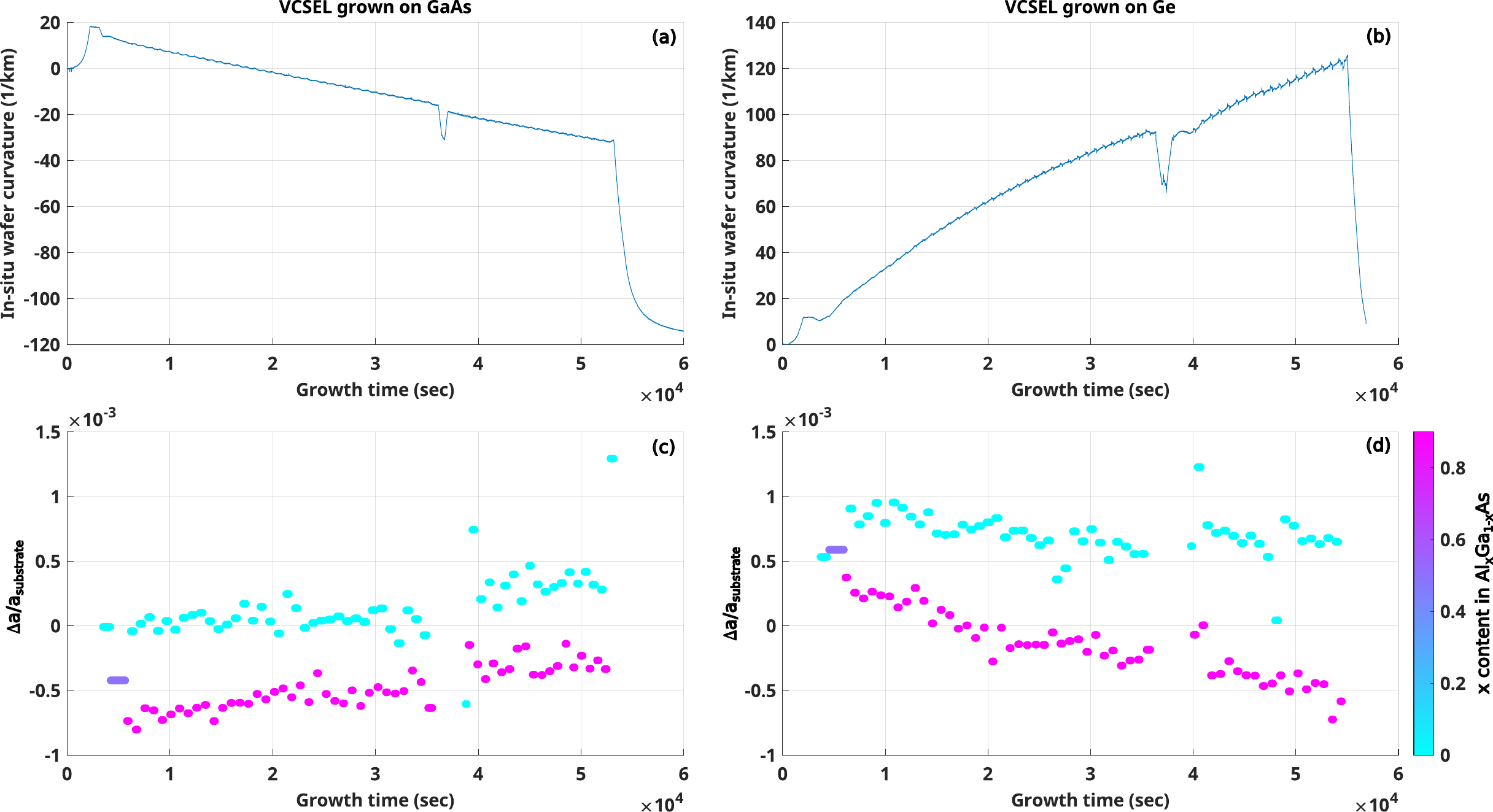}
  \caption{In-situ wafer curvature using the MIC method during the complete MBE growth of the VCSEL structure on GaAs substrate (a) and on germanium (b). Corresponding lattice mismatch calculated with the Stoney law for each individual VCSEL layers grown on GaAs (c) and Ge (d). The datapoints color corresponds to its Al concentration in AlGaAs alloy.}
  \label{fig:curvature}
\end{figure*}

Figures~\ref{fig:curvature} (a) and (b) show the in-situ time evolution of the wafer curvature during the complete epitaxial growth of VCSEL recorded by MIC, respectively, for a 940-nm VCSEL grown on GaAs substrate, taken as reference, and another on Ge substrate.
In both cases, the initial curvature was set to zero to make it easier to compare the two curves. 

The growth sequence begins with a substrate heatup from 100 \,\textdegree{}C to 720 \,\textdegree{}C under As overpressure for oxide desorption [$0-\sim 3000\,s$], during which no material is deposited. In both cases, a parabola-like curvature increase followed by a plateau is observed in this phase, primarily attributed to a temperature gradient across the wafer thickness\cite{maassdorf2013} and differential thermal expansion between the Ge substrate and the MOVPE GaAs buffer layer in the second case. After lowering the temperature to 680\textdegree{}C, decreasing the curvature of about $3-4\,km^{-1}$, a 100-nm GaAs layer is grown, followed by a 500-nm Al$_{0.5}$Ga$_{0.5}$As layer, which corresponds to the close lattice-matching condition with Ge at room temperature.
The following VCSEL growth sequence is composed of the bottom DBR [$\sim 5000-36000\,s$], the optical cavity [$\sim 36000-37000\,s$], and the top DBR [$\sim 37000-53000\,s$]. The last section corresponding to the large curvature reduction is due to the wafer cooling phase down to 100\textdegree{}C.
The difference in curvature between the initial and final stages shows the extent to which the wafer has bowed due to the epitaxy of the VCSEL structure, with GaAs showing very significant bowing and Ge showing limited bowing.

The main differences in the time evolution of the curvatures between the identical VCSEL structures grown respectively on GaAs and on Ge are, first, the opposite slope sign, and second, the overall linear variation on GaAs $vs$ nonlinear variation for the Ge case.
These behaviours can be explained by the fact that the lattices of GaAs, AlGaAs, and Ge change differently with temperature \cite{Madelung2004}. In particular, in the 500-700\textdegree{}C range, $a_{Ge} > a_{AlGaAs} > a_{GaAs}$, which explains the different signs of the derivative of curvature over time. In the case of growth on Ge, this implies that the balance of strains between GaAs and AlAs (or Al-rich AlGaAs) differs significantly between room temperature, where stress compensation occurs, and high temperature, where strains from GaAs and AlGaAs build up cumulatively.   

The lower part of Figures~\ref{fig:curvature} ((c) and (d)) are depicting the real-time evolutions of the lattice mismatch to the substrate, computed from the above in-situ curvature measurements at growth temperature, calculated with the Stoney law and given the wafer thickness \cite{BenSaddik2025}. Each coloured point corresponds to the individual thickest layers composing the VCSEL structure. The average $\Delta a/a$ for each AlGaAs composition correspond to the expected values at growth temperature, and considering the additional strain due to carbon doping for the section corresponding to the upper DBR. 
Here again, and in accordance with the curvature curves described above, the time progression in the case of growth on GaAs and Ge are very different. While the lattice mismatch between the GaAs and AlGaAs layers remains relatively constant throughout growth on a GaAs substrate, an increasingly large mismatch between these same layers is observed when the VCSEL is grown on Ge.
Although the variation in lattice mismatch for a given alloy composition over time may be due to flux drift, as appears to be the case for Al$_{0.9}$Ga$_{0.1}$As on GaAs (fig. ~\ref{fig:curvature}(c)), the more pronounced increasing gap through the VCSEL growth between the lattice mismatches of GaAs and AlGaAs on Ge substrate (fig. ~\ref{fig:curvature}(d)) could be explained by partial relaxation due to the accumulation of compressive strain. This would explain the ‘abnormal’ curved shape of the wafer curvature evolution visible during the growth of DBRs on Ge. 


\begin{figure}[h!]
  \centering
   \includegraphics[width=0.45\textwidth]{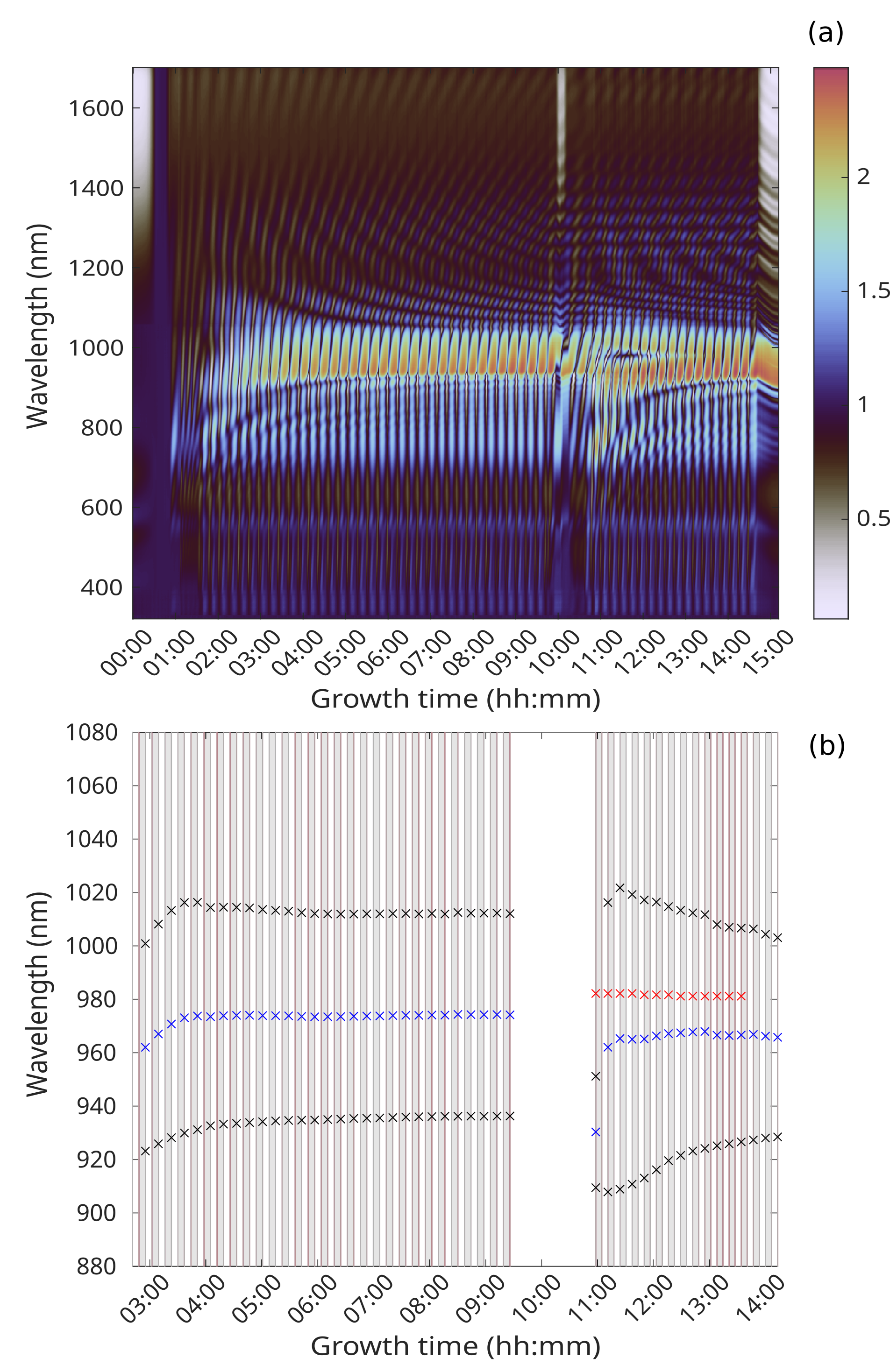}
   \caption{(a) In situ wafer reflectometry during the complete epitaxial growth of a VCSEL structure by MBE. (b) real-time detected characteristic VCSEL wavelengths during growth: DBR sidelobes edges (black crosses), DBR center wavelength (blue), Fabry-perot cavity dip (red). The alternating layers with high index contrast (GaAs/AlGaAs) are illustrated in grey and white shading.}
  \label{fig:Reflectance}
\end{figure}

Figure~\ref{fig:Reflectance}(a) displays the time resolved reflectance heatmap acquired over the 320-1700~nm spectral range. During the early stages of growth, distinct periodic interference fringes were observed across the spectrum, originating from reflections at the evolving surface of the initial buffer layers GaAs and Al$_{0.5}$Ga$_{0.5}$As. 

Figure~\ref{fig:Reflectance}(b) shows the characteristic wavelengths of the VCSEL structure deduced from this spectral measurement at each end of the high index layer (GaAs): the spectral position of the bottom and top DBRs, and the wavelength of the microcavity Fabry-Pérot resonance (FP). The DBR stopband could not be detected in the first few periods because the reflectivity was too low. Also, the FP dip cannot be detected in the uppermost part of the VCSEL structure due to the increasingly narrow width when stacking layers of the upper mirror.

Starting around 01:30, the bottom DBR starts, showing the building-up of the high reflectivity stopband centered at 974 nm (corresponding to about 935-936 nm at room-temperature based on past experience and estimated substrate temperature). 
Throughout the entire bottom DBR growth process, no wavelength drift of the high reflectivity stopband is observed (Fig. ~\ref{fig:Reflectance}(b)). Moreover, increasing edge steepness of the stopband edges, and regularly wavelength-spaced sidelobes on either side are visible, resulting in perfectly matched quarter-wavelength Bragg layers and consequently prevailing high reflectivity upon returning to room temperature.

Following the growth of the optical cavity, including the MQW gain region, between $09:56$ an $10:18$, the growth of the top DBR proceeds, revealing after few periods the microcavity resonance wavelength (dip) (Fig. ~\ref{fig:Reflectance}(b)). The resonance dip remains apparent and constant around 981-982\,nm (corresponding to 943 nm at room temperature), fading as the upper DBR grows as described before. 
At the end of the VCSEL growth, the reflectivity stopband is still observed and closely centered at the same wavelength as the previously seen bottom DBR and the cavity resonance dip.  


\begin{figure}[h!]
  \centering
  \includegraphics[width=0.45\textwidth]{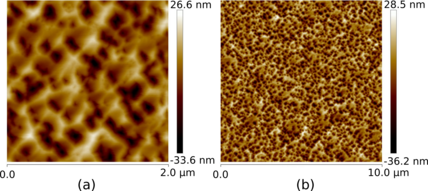}
  \caption{Atomic force microscopy (AFM) images of VCSEL epitaxial samples surface grown on Ge, with scanning areas of (a) $2.0 \times 2.0\,\mu m^2$ ($R_q=8.56\,nm$) and (b) $10 \times 10\,\mu m^2$ ($R_q=9.67\,nm$). }
  \label{fig:AFM}
\end{figure}

The $2\,\mu\text{m} \times 2\,\mu\text{m}$ Atomic force microscopy (AFM) scan (Figure~\ref{fig:AFM}a), acquired in tapping mode using a Bruker Dimension Icon system, shows a continuous surface with nanoscale height modulations and localized depressions. No cracks or step bunching are observed within this field of view. The surface does not exhibit a clear preferential lateral orientation. The root mean square roughness ($R_q$) is 8.56 nm, and the arithmetic average roughness ($R_a$) is 6.73 nm. The developed surface area exceeds the projected area by 3.97\%, and the peak-to-valley height range is approximately 60 nm. Although individual terraces are not resolved and the absence of large-scale discontinuities suggests a laterally uniform topography at the micron scale.

The $10\,\mu\text{m} \times 10\,\mu\text{m}$ scan (Figure~\ref{fig:AFM}b) reveals a similar surface texture with $R_q = 9.67\,nm$, $R_a = 7.75\,nm$, and a surface area difference of 2.17\%. The slightly increased roughness may be influenced by the scan size or image processing parameters. No long-range step bunching or cracks are evident, indicating that the surface remains laterally continuous over larger areas. AFM provides no subsurface information, but the measured topography indicates that the uppermost layers were grown without surface-breaking defects or morphological instabilities. The measured surface roughness values are high compared to a VCSEL grown on GaAs. This roughness could be improved by optimising the growth conditions, particularly the temperature. Furthermore, it does not appear to affect significantly the high reflectivity of Bragg mirrors, as shown by the following FTIR measurements.

\begin{figure}[h!]
  \centering
  \includegraphics[width=0.45\textwidth]{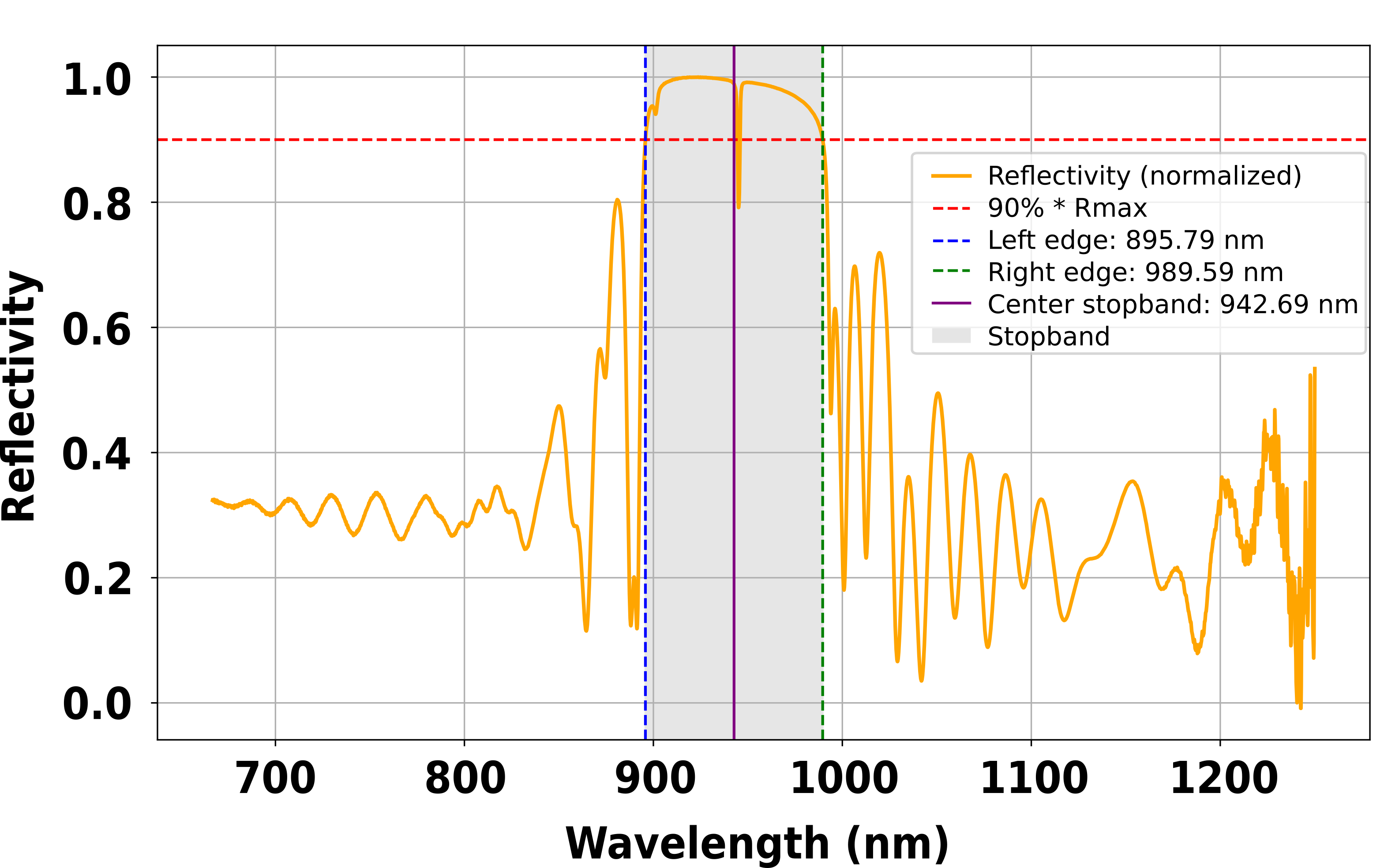}
  \caption{Normal incidence reflectivity spectrum showing a stopband centered at 942.7\,nm with a width of 93.6\,nm. This confirms accurate optical thickness control in the DBR stack.}
  \label{fig:reflectance}
\end{figure}

Complementary optical characterization was performed ex situ by measuring normal incidence reflectivity spectrum. As shown in Figure~\ref{fig:reflectance}, a high-reflectivity wide stopband is observed, centered at 942.7\,nm with a full width at 90\% reflectivity of 93.6\,nm, spanning from 895.8\,nm to 989.5\,nm. This spectral response matches the design target and closely aligns with the final in situ reflectometry spectrum, as discussed previously. The steep spectral transitions and broad high reflectivity plateau indicate minimal optical scattering and a strong refractive index contrast within the DBR layers.

This confirms the good structural and optical properties of the MBE-grown VCSEL structure, validating the process as suitable for efficient and narrow linewidth microcavity emitters integrated on Ge substrates.

\begin{figure}[t]
    \centering
    \includegraphics[width=\columnwidth]{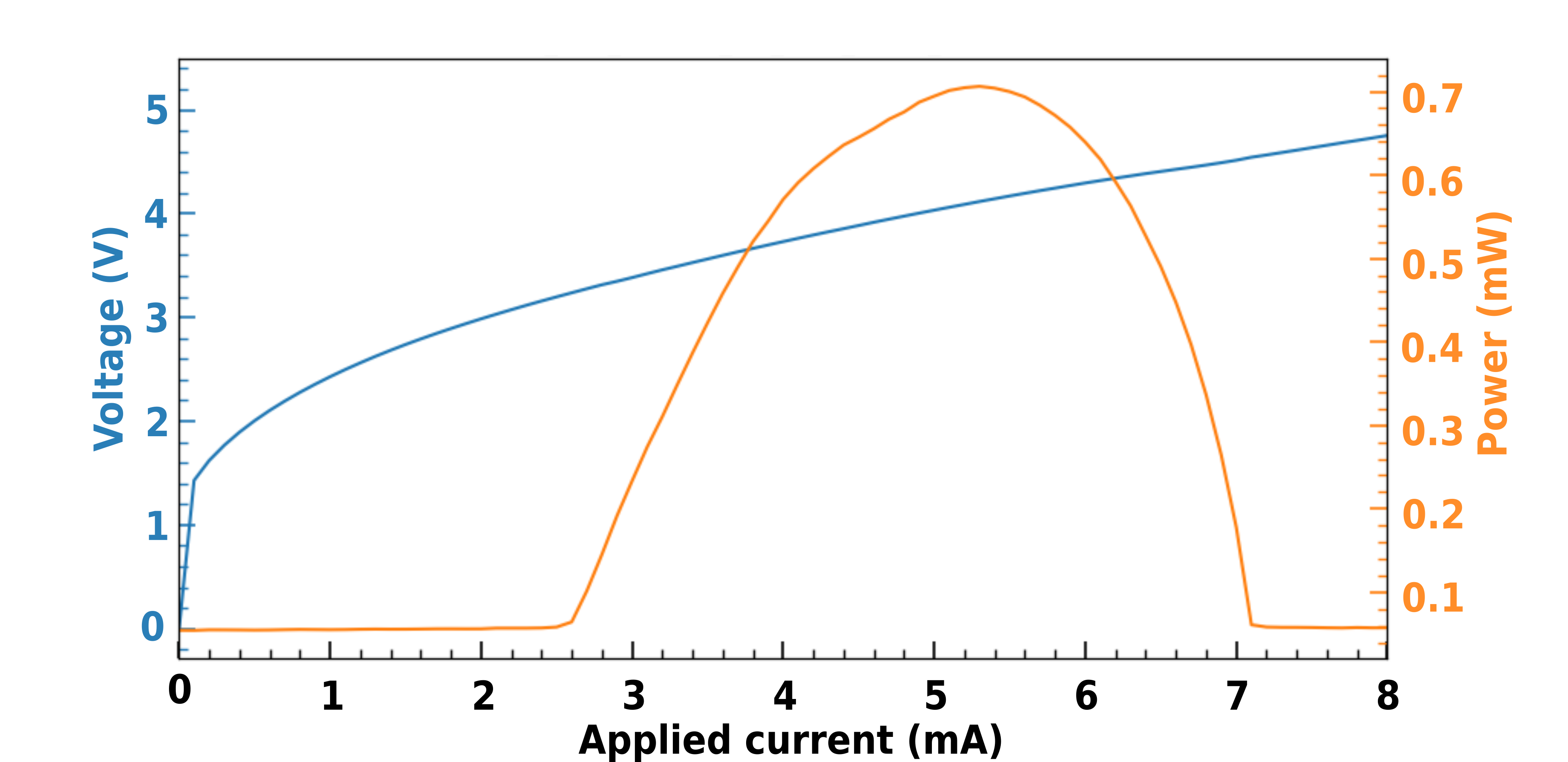}
    \caption{
    LIV characteristic of a 35~µm-diameter VCSEL measured under continuous-wave electrical injection at 25\,°C.}
    \label{fig:liv35um}
\end{figure}

The lasing behavior of the VCSELs was evaluated under continuous-wave electrical injection at 25~ºC. Figure~\ref{fig:liv35um} shows a representative LIV characteristic measured on a device with a 35~µm mesa diameter with an estimated $11\,\mu m$ oxide aperture. The voltage exhibits the expected diode-like increase at low current and reaches approximately 3.1~V at the lasing threshold. A clear onset of stimulated emission occurs at a threshold current of $I_\mathrm{th} \approx 2.8$~mA, after which the optical output rises sharply. The device reaches a maximum emitted power of $\sim 0.7$~mW at a drive current of about 5.2~mA. Beyond this point, a pronounced thermal rollover is observed, with the output power decreasing as current increases further, a behavior attributed to self-heating in the cavity and limited heat extraction through the Ge substrate.

The measured performance confirms that the MBE-grown active region and DBR mirrors provide efficient vertical optical confinement and low-threshold lasing on Ge(001). At the same time, the observed rollover highlights the need for optimized thermal management strategies, including improved heat sinking or modified mirror doping and thickness to reduce series resistance and internal heating.

In conclusion, this work reports on the epitaxy of AlGaAs-based VCSELs by MBE on germanium wafer. In-situ curvature and reflectometry measurements performed in real time during growth enabled detailed analysis of the mechanical and optical properties of the structure during MBE growth. The curvature measurement related to internal stresses revealed an abnormal evolution of epitaxial strains during the growth of VCSELs on Ge, which could be explained by a strain relaxation between AlGaAs and Ge at the growth temperature. However, this hypothesis remains to be verified by other analysis methods. Despite this atypical behaviour for the VCSEL growth on Ge, the optical properties are similar to those of VCSEL structures grown on GaAs. Finally, VCSEL devices grown by MBE on Ge show low threshold laser performance at room temperature and under continuous electrical injection. 
These findings represent a major breakthrough in the development of MBE-grown VCSELs on germanium substrates, paving the way for their integration into high-performance photonic transmission systems and advanced sensing applications.

\begin{acknowledgments}
The authors acknowledge the financial support of the EC Horizon PhotoGeNIC project (GA 101069490), and the Polish National Centre for Research and Development (MAZOWSZE/0032/19-00). This work was supported by the LAAS-CNRS micro and nanotechnologies platform, a member of the French RENATECH network, and by the joint laboratory EPICENTRE between CNRS-LAAS and RIBER.
\end{acknowledgments}

\section*{Author Declarations}
The authors have no conflicts to disclose.



\appendix

\nocite{*}
\bibliography{aipsamp}

\end{document}